\documentclass[reprint,superscriptaddress,aps]{revtex4-1}
\usepackage{amsmath}
\usepackage{amssymb}
\usepackage{bm}
\usepackage{braket}
\usepackage{color}
\usepackage[varg]{txfonts}
\usepackage{varwidth}
\usepackage{dcolumn}
\usepackage[breaklinks,colorlinks=true,linkcolor=blue,urlcolor=cyan,citecolor=blue]{hyperref}
\usepackage{graphicx}

\begin{document}

\title{Nematicity and fractional magnetization plateaus induced by spin-lattice coupling in the classical kagome-lattice Heisenberg antiferromagnet}

\author{Masaki Gen}
\affiliation{Department of Advanced Materials Science, University of Tokyo, Kashiwa 277-8561, Japan}

\author{Hidemaro Suwa}
\email{suwamaro@phys.s.u-tokyo.ac.jp}
\affiliation{Department of Physics, The University of Tokyo, Tokyo 113-0033, Japan}

\begin{abstract}

We investigate the effect of spin-lattice coupling (SLC) on the magnetic properties of the classical kagome-lattice Heisenberg antiferromagnet (KHAF) using improved Monte Carlo updates.
The lattice modes are represented by Einstein site phonons, which introduce effective further-neighbor spin interactions in addition to the nearest-neighbor biquadratic interactions.
In the weak SLC, the macroscopically degenerate coplanar ground state remains at zero field, while a $\sqrt{3} \times \sqrt{3}$ ordered phase accompanied by a 1/3-magnetization plateau appears in external magnetic fields.
In the strong SLC, we find a nematic order at zero field and a 1/9-magnetization plateau associated with a $3 \times 3$ collinear order.
Near the phase transition between the 1/9- and 1/3-plateau states, the ergodicity in the single spin flip is practically broken, and slow dynamics appear.
We propose that relevant KHAFs with strong SLC would be realized in spinel-based materials.

\end{abstract}

\date{\today}
\maketitle
\section{Introduction}
Over the past three decades, the kagome-lattice Heisenberg antiferromagnet (KHAF) has been a central playground for exploring exotic magnetic states introduced by geometrical frustration.
The ground state of the classical KHAF is infinitely degenerate, while thermal fluctuations partially lift it and favor a disordered coplanar spin state \cite{1992_Cha, 1993_Rei, 2008_Zhi}.
In the spin-1/2 case, the ground state is believed to be a quantum spin liquid, the nature of which has been actively discussed \cite{2007_Ran, 2008_Jia, 2011_Yan, 2011_Iqb, 2012_Dep, 2013_Nis, 2017_Lia, 2017_Mei, 2018_Che}.
For both cases, even small perturbations, such as the Dzyaloshinskii-Moriya (DM) interaction \cite{2002_Elh, 2008_Cep, 2010_Mes} and further-neighbor (FN) interactions \cite{2011_Mes, 2012_Mes, 2015_Gon, 2015_Kol}, can induce various magnetic long-range orders (LROs).

Also in external magnetic fields, the KHAF can exhibit rich magnetic phases dressed with fractional magnetization plateaus.
The typical one is a 1/3-magnetization plateau induced by the quantum effect for arbitrary spin values \cite{2001_Hid, 2011_Sak, 2015_Pic, 2018_Nak, 2002_Sch, 2013_Cap, 2016_Pic, 2018_Pla, 2020_Sch}.
Even in the classical limit, a 1/3-magnetization plateau with a collinear spin-liquid state is stabilized by thermal fluctuations due to the order-by-disorder effect \cite{2002_Zhi, 2011_Gvo}.
Of particular interest is a series of magnon crystals localized on hexagons of the kagome lattice \cite{2013_Nis, 2002_Sch, 2013_Cap, 2016_Pic, 2018_Pla, 2020_Sch}, which was impressively evidenced by multi-step magnetization jumps observed in Cd-kapellasite \cite{2019_Oku}.
Furthermore, a 1/9-magnetization plateau has been predicted for spin-1/2 by the density matrix renormalization group method \cite{2013_Nis} and tensor network algorithms \cite{2018_Che, 2016_Pic, 2019_Oku}.
However, exact diagonalization studies have challenged the presence of it \cite{2011_Sak, 2018_Nak}.
There has been no experimental evidence of the 1/9-magnetization plateau in Cd-kapellasite and herbertsmithite \cite{2019_Oku, 2020_Oku}, which are the most likely materials for the ideal spin-1/2  KHAF \cite{2009_Nyt, 2005_Sho}.

Most previous theoretical works on the KHAF did not take the phonon contribution into account.
In highly frustrated magnets, spin-lattice coupling (SLC) often plays an essential role in the determination of the magnetic state: e.g., the zero-field zigzag order and the 1/5-magnetization plateau in a triangular-lattice antiferromagnet CuFeO$_{2}$ \cite{2000_Mit, 2006_Ye}, the 2-up–2-down N\'{e}el order, and the 1/2-magnetization plateau in pyrochlore-based chromium spinels \cite{2006_Ued, 2007_Mat, 2008_Koj, 2009_Ji, 2010_Mat, 2015_Kim, 2019_Gen, 2020_Gen}, and more complex field-induced phases in a ferrimagnetic spinel MnCr$_{2}$S$_{4}$ \cite{2017_Tsu, 2020_Miy,2021_Yam}.
It has been revealed that the microscopic magnetoelastic theories assuming local bond-phonon \cite{2020_Miy, 2004_Pen, 2010_Sha,2021_Aoy} and site-phonon modes \cite{2020_Gen, 2021_Aoy, 2006_Ber, 2016_Aoy, 2008_Wan} successfully account for these SLC-induced LROs.
For the kagome lattice, however, the theoretical investigation on the effect of SLC exists only for the spin ice model \cite{2013_Alb}, whereas one for the Heisenberg model is lacking apart from a brief remark found in Ref.~\cite{2008_Wan}.

\begin{figure}[t]
\centering
\includegraphics[width=\linewidth]{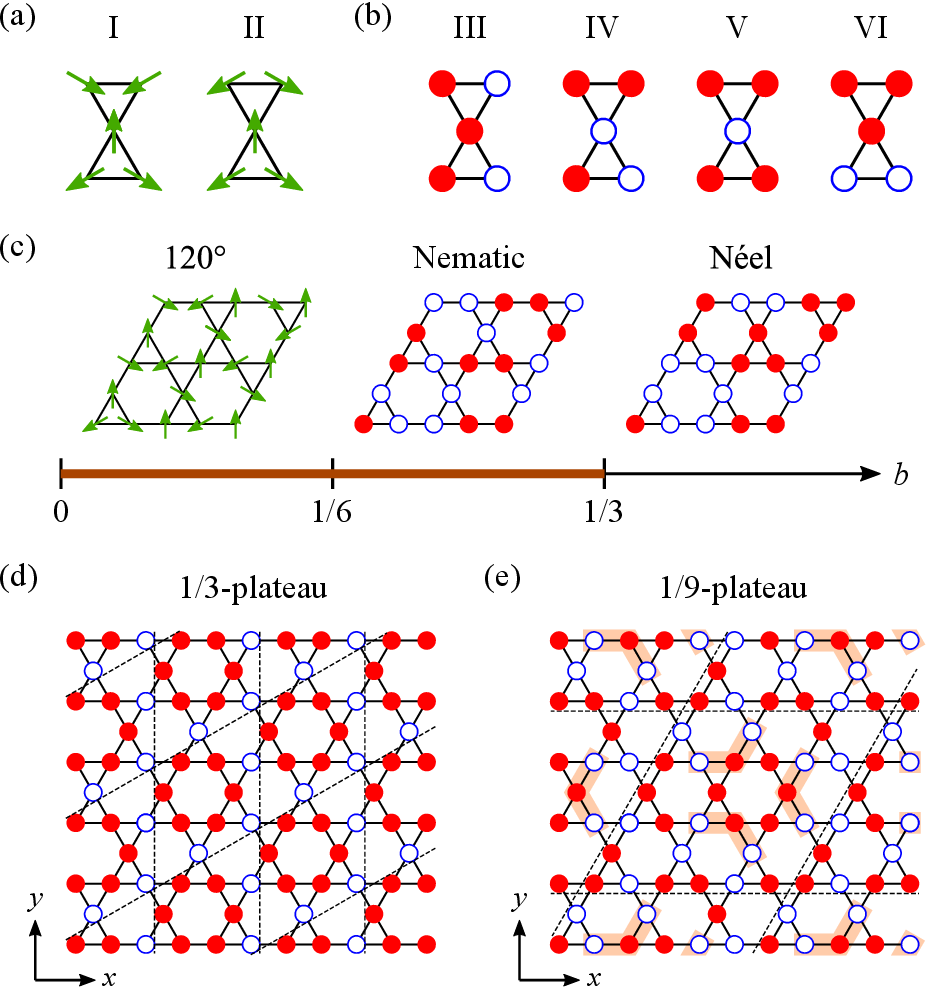}
\caption{(a) Local spin configurations of the 120$^{\circ}$ coplanar state characterized by two corner-sharing triangles with (I) the identical chirality and (II) the opposite chirality. (b) Four relevant local spin configurations of the observed collinear states. (c) Ground-state phase diagram at zero field with respect to the SLC parameter $b$. Twelve-sublattice $2 \times 2$ LRO exists in the ``N\'{e}el" phase, whereas macroscopic degeneracy remains in the 120$^{\circ}$ coplanar and the nematic phases. Typical spin configurations in these degenerate phases are illustrated. (d) Magnetic structure of the 1/3-magnetization plateau state. (e) Magnetic structure of the 1/9-magnetization plateau state. In (b)–(e), the red solid (blue open) circles represent up (down) spins. In (d) and (e), the magnetic unit cell is illustrated by a parallelogram background. Flipping the spins of the orange clusters shown in (e) connects the two plateau states.}
\label{Fig1}
\end{figure}

In this paper, we reveal comprehensive phase diagrams of the classical KHAF coupled to local site-phonon modes, using the Monte Carlo (MC) method and advanced sampling techniques.
In the weak SLC, the 120$^{\circ}$ coplanar ground state at zero field remains, while a sequence of field-induced phase transitions takes place, exhibiting a 1/3-magnetization plateau associated with a $\sqrt{3} \times \sqrt{3}$ collinear order [Fig.~\ref{Fig1}(d)].
In the strong but physically reasonable SLC, we find a nematically ordered ground state at zero field and a novel 1/9-magnetization plateau associated with a $3 \times 3$ collinear order [Fig.~\ref{Fig1}(e)] in the low-field region.

\section{Model}
\label{sec:model}
We consider the spin Hamiltonian of the KHAF with the elastic and the Zeeman terms:
\begin{equation}
\label{H}
{\mathcal{H}}=\sum_{\langle i, j\rangle} J_{ij} {\mathbf S}_{i} \cdot {\mathbf S}_{j}+\frac{c}{2}\sum_{i}{|{\mathbf u}_{i}|^{2}}-h\sum_{i}S_{i}^{z},
\vspace{-0.1cm}
\end{equation}
where $\langle i, j\rangle$ runs over all the nearest-neighbor (NN) sites, $J_{ij}~(>0)$ is the antiferromagnetic exchange coupling, ${\mathbf S}_{i}$ is the classical spin at site $i$ normalized to $|{\mathbf S}_{i}|=1$, $c~(>0)$ is the spring constant, ${\mathbf u}_{i}$ is the displacement at site $i$ from its original position ${\mathbf r}_{i}^{0}$, and $h$ is the strength of an external magnetic field applied along the $z$ axis.
The exchange striction is introduced assuming $J_{ij}$ linearly modulated by the bond-length change provided that $|{\mathbf u}_{i}|/|{\mathbf r}_{i}^{0}| \ll 1$: $J_{ij} \equiv J(|{\mathbf r}_{ij}^{0}+{\mathbf u}_{i}-{\mathbf u}_{j}|) \approx J+(dJ/dr)|_{r=|{\mathbf r}_{ij}^{0}|}{\mathbf e}_{ij} \cdot ({\mathbf u}_{i}-{\mathbf u}_{j})$, where $J \equiv J(|{\mathbf r}_{ij}^0|)$, ${\mathbf r}_{ij}^{0} \equiv {\mathbf r}_{i}^{0}-{\mathbf r}_{j}^{0}$, and ${\mathbf e}_{ij} \equiv {\mathbf r}_{ij}^{0}/|{\mathbf r}_{ij}^{0}|$.
We also assume $J$ and $(dJ/dr)|_{r=|{\mathbf r}_{ij}^{0}|}$ independent of the site.
For the lattice degrees of freedom, the displacements ${\mathbf u}_{i}$ are independent of each other in the absence of the SLC, i.e., Einstein site phonons \cite{2006_Ber}.
The Hamiltonian is invariant under the simultaneous sign reversal of the derivative $(dJ/dr)|_{r=|{\mathbf r}_{ij}^{0}|}$ and the lattice displacements ${\mathbf u}_i$ for all $i$.
Thus, the sign of $(dJ/dr)|_{r=|{\mathbf r}_{ij}^{0}|}$ is irrelevant to the physics of this system.

In this model, the exchange coupling depends only on the distance in the direction of the interatomic vector.
This approximation is justified if $|{\mathbf u}_{i}|/|{\mathbf r}_{i}^{0}| \ll 1$.
The shift in the direction perpendicular to the interatomic vector changes the distance by $O(|{\mathbf u}_{i}|^2/|{\mathbf r}_{i}^{0}|)$, while the shift in the direction of the interatomic vector does by $O(|{\mathbf u}_{i}|)$.
The effect of the shift in the perpendicular direction should be negligible.

Let us next consider the effective spin model.
The Hamiltonian~(\ref{H}) can be expressed by
\begin{equation}
\label{H2}
{\mathcal{H}}=J\sum_{\langle i, j\rangle} {\mathbf S}_{i} \cdot {\mathbf S}_{j}+\frac{c}{2}\sum_{i}{|{\mathbf u}_{i} - \bar{\mathbf u}_i|^{2}}-\frac{c}{2}\sum_{i}{|\bar{\mathbf u}_{i}|^{2}}-h\sum_{i}S_{i}^{z},
\end{equation}
where
\begin{equation}
\bar{\mathbf u}_i = \sqrt{\frac{Jb}{c}} \sum_{j \in N(i)} {\mathbf e}_{ij} ({\mathbf S}_i \cdot {\mathbf S}_j)
\end{equation}
is the average displacement given the spin state.
We here adopt the canonical ensemble and exactly integrate out the lattice degrees of freedom ${\mathbf u}_{i}$ using the standard Gaussian integration \cite{2006_Ber}. The effective spin Hamiltonian is given by
\begin{equation}
\label{H_eff}
{\mathcal{H}_{\rm eff}}=J\sum_{\langle i, j\rangle}[{\mathbf S}_{i} \cdot {\mathbf S}_{j} - b({\mathbf S}_{i} \cdot {\mathbf S}_{j})^2]+{\mathcal{H}_{\rm FN}}-h\sum_{i}S_{i}^{z},\\
\end{equation}
\vspace{-0.5cm}
\begin{equation}
\label{H_FN}
{\mathcal{H}_{\rm FN}}=-\frac{Jb}{2}\sum_{j\neq k\in N(i)}{\mathbf e}_{ij} \cdot {\mathbf e}_{ik}({\mathbf S}_{i} \cdot {\mathbf S}_{j})({\mathbf S}_{i} \cdot {\mathbf S}_{k}),
\end{equation}
where the dimensionless parameter $b$ represents the strength of the SLC defined by $b \equiv (1/cJ)[(dJ/dr)|_{r=|{\mathbf r}_{ij}^{0}|}]^{2}$~($>$0), and $N(i)$ is the set of the NN sites of site $i$.
Evidently, the SLC produces the biquadratic terms and the three-body quartic terms ${\mathcal{H}_{\rm FN}}$.
The former favors collinear spin configurations, while the latter acts as effective FN interactions.
The energies of ${\mathcal{H}_{\rm FN}}$ for several local spin configurations appearing in the 120$^{\circ}$ coplanar state and the relevant collinear states are $E^{\rm (I)}=E^{\rm (II)}=Jb/2$, $E^{\rm (III)}=E^{\rm (IV)}=0$, $E^{\rm (V)}=2Jb$, and $E^{\rm (VI)}=-4Jb$, where the supersubscripts represent the corresponding spin configurations shown in Figs.~\ref{Fig1}(a) and \ref{Fig1}(b).

To study the thermodynamic properties of this system, we performed classical MC simulations for $N=3 L^{2}$ sites up to $L=72$ with periodic boundaries.
Combining the replica-exchange MC \cite{1996_Huk}, we introduce a microcanonical update \cite{1981_Adl, 1996_Alo} and a multi-spin-flip in addition to the conventional updates.
Our MC update is more than 40 times as efficient as the previously proposed ones \cite{2014_Shi, 2021_Aoy}.
Readers are referred to the Appendices for details of our simulations.

\section{Zero-field case}
\label{sec:zero-h}
The ground-state phase diagram at zero field with respect to the SLC parameter $b$ is shown in Fig.~\ref{Fig1}(c) (also presented in Ref.~\citenum{2008_Wan}).
To obtain the phase diagram, we first ran MC simulations at low enough temperatures varying $b$ and confirmed that the ground state changed from the 120$^\circ$ coplanar state to collinear states.
We then considered all the local collinear configurations and calculated their energies.
The ground-state phase diagram was obtained by minimizing the total energy among all the collinear states and comparing it to the energy of the 120$^\circ$ coplanar state.

In the weak SLC region ($b < 1/6$), the ground state is the 120$^{\circ}$ coplanar state, which satisfies the local condition of $\sum_{i \in \bigtriangleup}{\mathbf S}_{i}=0$, where $\bigtriangleup$ denotes the triangular unit of the kagome lattice.
The macroscopic degeneracy remains in this state because the spin configurations I and II [Fig.~\ref{Fig1}(a)] have the same energy even with the SLC.
In the strong SLC region ($b > 1/6$), on the other hand, collinear states become stable due to the dominant biquadratic terms, breaking the local condition of $\sum_{i \in \bigtriangleup}{\mathbf S}_{i}=0$.
For $1/6 < b < 1/3$, the ground state is nematically ordered, in which the spin configurations with the same energy, III, IV, and their flipped (up $\leftrightarrow$ down) counterparts, are randomly arranged.
For $b > 1/3$, the ground state has a 12-sublattice $2 \times 2$ LRO, in which the spin configurations III and VI are regularly arranged in the same ratio.

\begin{figure}[t]
\centering
\includegraphics[width=\linewidth]{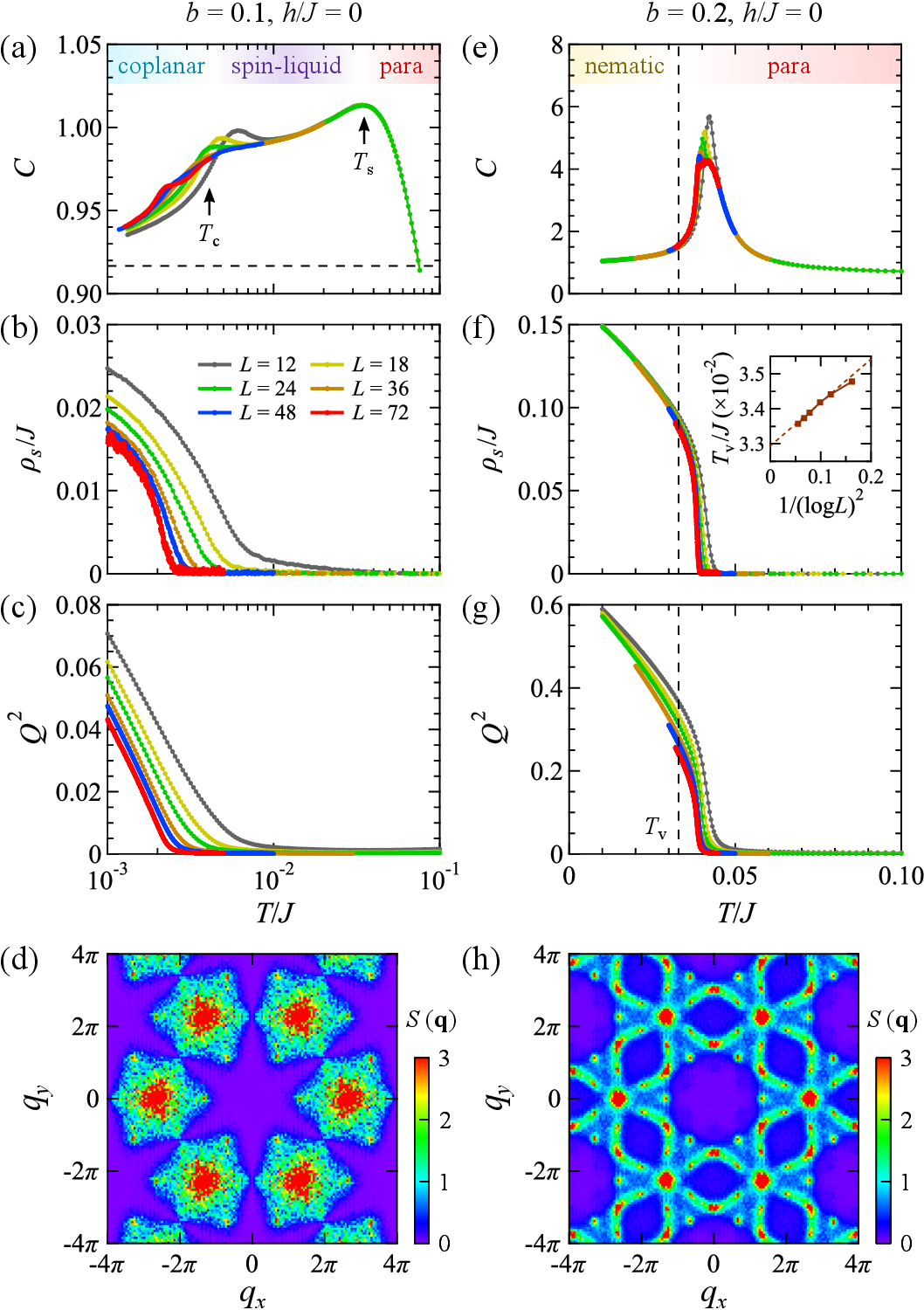}
\caption{(a)(e) Specific heat $C$; (b),(f) spin stiffness $\rho_{s}$; and (c),(g) nematic order parameter $Q^{2}$ as a function of temperature at zero field for $b=0.1$ [(a)–(c)] and $b=0.2$ [(e)–(g)]. The dashed line in (a) indicates the expected value $11/12$ in $T/J \to 0$. The inset of (f) shows the finite-size scaling of the vortex-binding transition point (see text for details). (d),(h) Spin structure factor for $L=36$ and $b=0.1$ at $T/J=0.001$ (d) and for $b=0.2$ at $T/J=0.02$ (h). The ground state is the 120$^\circ$ coplanar spin state for $b=0.1$ and the nematically ordered state for $b=0.2$.}
\label{Fig2}
\end{figure}

Figure~\ref{Fig2} shows the temperature dependence of the specific heat $C$, the spin stiffness $\rho_s$, and the nematic order parameter $Q^2$ at zero field for $b=0.1$ [(a)–(c)] and $b=0.2$ [(e)–(g)]. 
Figures~\ref{Fig2}(d) and \ref{Fig2}(h) show the spin structure factor $S(\mathbf{q})$ at $T/J=0.001$ for $b=0.1$ and at $T/J=0.02$ for $b=0.2$.
Each physical quantity can be calculated by
\begin{equation}
\label{C}
C=\frac{\langle {\mathcal{H}_{\rm eff}}^2 \rangle - \langle {\mathcal{H}_{\rm eff}} \rangle^2}{NT^2}=\frac{1}{N}\frac{d \langle {\mathcal{H}_{\rm eff}} \rangle}{dT},
\end{equation}
\vspace{-0.3cm}
\begin{eqnarray}
\label{rho_s}
\rho_s &=& - \frac{\sqrt{3}}{4N} \sum_{\langle i, j \rangle} \left\langle J_{ij} \left( S^x_i S^x_j + S^y_i S^y_j \right) \right\rangle \nonumber \\
&&- \frac{2 \sqrt{3}}{NT} \left\langle \left[ \sum_{\langle i,j \rangle} J_{ij} \left(S^x_i S^y_j - S^y_i S^x_j \right) {\mathbf e} \cdot {\mathbf r}^0_{ij} \right]^2 \right\rangle,
\end{eqnarray}
\vspace{-0.2cm}
\begin{equation}
\label{Q2}
Q^2=\frac{1}{N^2} \sum_{\langle i, j\rangle} \left\langle ({\mathbf S}_i \cdot {\mathbf S}_j)^2 \right\rangle- \frac{1}{3},
\end{equation}
\vspace{-0.3cm}
\begin{equation}
\label{Sq}
S(\mathbf{q}) = \frac{1}{N} \left\langle \left| \sum_j \mathbf{S}_j e^{i \mathbf{q} \cdot \mathbf{r}^0_j} \right|^2 \right\rangle,
\end{equation}
where ${\mathbf e}$ is a unit twist vector, whose direction is arbitrary in the two-dimensional (2D) system \cite{2011_Gvo}, and $J_{ij}$ depends on local lattice displacements.
The specific heat was obtained by computing ${d \langle {\mathcal{H}_{\rm eff}} \rangle}/{dT}$ numerically.

For $b=0.1$, the specific heat exhibits two broad peaks around two crossover temperatures $T_{\rm s}/J \approx 0.04$ and $T_{\rm c}/J \approx 0.004$ [Fig.~\ref{Fig2}(a)], as in the classical KHAF without the SLC \cite{2008_Zhi}.
The spin-liquid state satisfying the local condition of $\sum_{i \in \bigtriangleup}{\mathbf S}_{i}=0$ appears below $T_{\rm s}$, and the coplanar state is further selected below $T_{\rm c}$ due to the additional zero modes \cite{1992_Cha}.
The correlation length of the magnetic and nematic orders exponentially diverges in $T \to 0$.
In finite-size systems, $\rho_s$ and $Q^2$ become nonzero at a temperature such that the correlation length reaches the system length.
The ordering temperature caused by the finite-size effect logarithmically decreases as $L$ increases, as shown in Figs.~\ref{Fig2}(b) and \ref{Fig2}(c).
As seen in Fig.~\ref{Fig2}(a), a shoulder appears in the specific heat at this ordering temperature, expected to disappear in the thermodynamic limit.
We also confirmed that the $\sqrt{3} \times \sqrt{3}$ magnetic order starts to develop at the same temperature (not shown).
The spin structure factor, shown in Fig.~\ref{Fig2}(d), is consistent with the 120$^\circ$ coplanar state \cite{2008_Zhi}.
These features imply that the zero-field magnetism inherent in the classical KHAF is robust to the weak SLC.

For $b=0.2$, the specific heat exhibits a prominent peak around $T/J \approx 0.04$ [Fig.~\ref{Fig2}(e)].
The spin state at low temperatures breaks the local condition of $\sum_{i \in \bigtriangleup}{\mathbf S}_{i}=0$ and has a nematic order, as shown in Fig.~\ref{Fig2}(g).
The spin structure factor, shown in Fig.~\ref{Fig2}(h), exhibits the characteristic structure of the nematically ordered state in the kagome lattice.
The order-parameter space of the nematic order is the real projective plane $\mathbb{R}P^{2}$, and its first homotopy group is $\pi_{1}\left(\mathbb{R}P^{2}\right)=\mathbb{Z}_{2}$.
Thus, a $\mathbb{Z}_2$ vortex emerges from the nematically ordered state as a point defect, called a disclination point.

In analogy to the Berezinskii-Kosterlitz-Thouless (BKT) transition, the topological transition at which $\mathbb{Z}_2$ vortex pairs are bound has been arguably discussed \cite{1984_Kaw, 2010_Kaw, 2010_Oku, 2020_Bon}.
Numerical studies showed that several nematic models exhibited universal scaling functions \cite{2020_Bon}.
However, renormalization group analyses pointed out that the fixed point dictated by the $\mathbb{Z}_2$ vortex binding might be out of the physical parameter space, resulting in a sharp crossover \cite{1996_Has, 1998_Cat}.
Whether it is a phase transition or a crossover, the spin correlation length $\xi$ becomes enormous at the transition and much longer than the numerically accessible system sizes, $\xi \sim 10^9$ for an $\mathbb{R}P^{2}$ model \cite{1996_Has}. This is consistent with the seemingly converging behavior of $\rho_s$ shown in Fig.~\ref{Fig2}(f) even though $\rho_s \to 0$ eventually in $L \to \infty$ at $T/J>0$.
As in the standard BKT transition, we estimate the transition or the crossover temperature $T_{\rm v}$ assuming the scaling $T_{\rm v}(L) - T_{\rm v}(\infty) \propto \frac{1}{(\ln L)^2}$, where $T_{\rm v}(L)$ satisfies the Nelson-Kosterlitz formula $\rho_s(L,T_{\rm v}(L))=\frac{2}{\pi v^2}T_{\rm v}(L)$ \cite{1977_Nel} for each $L$ with $v=1/2$ being the vorticity of the $\mathbb{Z}_2$ vortex.
This scaling should be valid for $1 \ll L \ll \xi$, and the extrapolation yields $T_{\rm v}/J=0.03294(8)$, as shown in the inset of Fig.~\ref{Fig2}(f).
Note that as in the case for $b=0.1$, a shoulder in the specific heat [Fig.~\ref{Fig2}(e)] is barely seen at the temperature at which the correlation length reaches the system length and expected to disappear in the thermodynamic limit.

\begin{figure}[t]
\centering
\includegraphics[width=\linewidth]{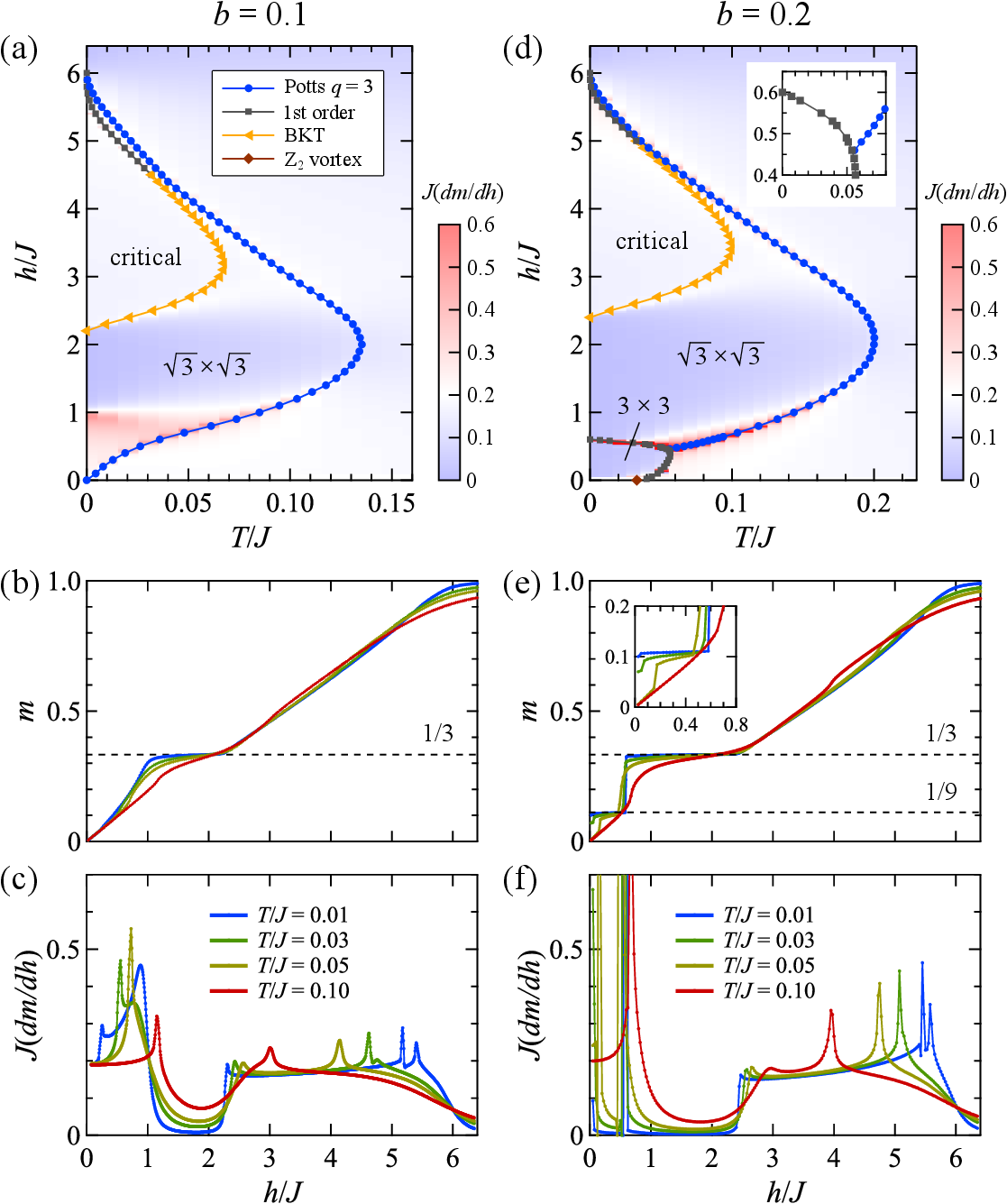}
\caption{Magnetic-field induced phase transitions for $b=0.1$ [(a)–(c)] and $b=0.2$ [(d)–(f)]: (a),(d) $h$–$T$ phase diagrams; (b),(e) magnetization curves; and (c),(f) their field derivatives for several temperatures. The contour map in (a) and (d) displays the value of $J(dm/dh)$. The inset of (d) shows an enlarged view of the phase boundary between the $\sqrt{3} \times \sqrt{3}$ and the $3 \times 3$ LRO phases. The inset of (e) shows an enlarged view of the 1/9-magnetization plateau.}
\label{Fig3}
\end{figure}

\section{In-field case}
\label{sec:in-h}
Next, we focus on the in-field properties.
Figures~\ref{Fig3}(a) and \ref{Fig3}(d) show the $h$–$T$ phase diagrams for $b=0.1$ and $b=0.2$, respectively.
The first-order transition points were extrapolated from the peak position of the specific heat, and the other transition points were estimated using the stochastic approximation \cite{2015_Yas}.
For $b=0.1$, the spins form a Y shape in low fields, a 2-up–1-down state in intermediate fields, and a V shape in high fields, with the $\sqrt{3} \times \sqrt{3}$ LRO [Fig.~\ref{Fig3}(a)], and a robust 1/3-magnetization plateau accompanies the 2-up–1-down state.
The magnetization $m$ and its field derivative $J(dm/dh)$ are plotted for several temperatures in Figs.~\ref{Fig3}(b) and \ref{Fig3}(c), respectively.
Figure~\ref{Fig1}(d) illustrates the threefold degenerate 1/3-plateau state, comprised of the spin configurations III and V in the ratio of 2:1 [Fig.~\ref{Fig1}(b)].
The phase transition to the $\sqrt{3} \times \sqrt{3}$ LRO phase is characterized by the 2D $q$-state Potts universality class with $q=3$ describing the $\mathbb{Z}_3$ symmetry breaking.
At higher fields above the 1/3-magnetization plateau, the BKT transition occurs in the spin $xy$ components, turning into the first-order transition near the saturation below $T/J \approx 0.03$.

The phase diagram for $b=0.2$ is qualitatively similar to that for $b=0.1$, except in the low-field region.
Notably, the macroscopically degenerate nematic state at zero field includes a 5-up–4-down state.
Thus, a 1/9-magnetization plateau appears under an infinitesimal magnetic field, found robust to thermal fluctuations.
Figure~\ref{Fig1}(e) illustrates the 18-fold degenerate 1/9-plateau state, which breaks a $\mathbb{Z}_3 \times \mathbb{Z}_6$ symmetry and possesses a $3 \times 3$ LRO, comprised of the spin configurations III and IV, and spin-flip (up $\leftrightarrow$ down) IV, in the ratio of 3:4:2 [Fig.~\ref{Fig1}(b)].
While the $\mathbb{Z}_3$ symmetry corresponds to the translation by $(1, 0)$ in units of the lattice constant, the $\mathbb{Z}_6=\mathbb{Z}_2 \times \mathbb{Z}_3$ symmetry consists of the $\mathbb{Z}_2$ for the inversion ($x \leftrightarrow -x$) and the $\mathbb{Z}_3$ for the translation by $(\frac{3}{2}, \frac{\sqrt{3}}{2})$.
In higher fields, the magnetization jumps from $m = 1/9$ to 1/3 [Fig.~\ref{Fig3}(e)].
The energy densities of the 1/9- and 1/3-plateau states are $E_{m=1/9}=-\frac{2}{3}J(1+3b)-\frac{h}{9}$ and $E_{m=1/3}=-\frac{2}{3}J(1+2b)-\frac{h}{3}$, respectively, so that the first-order transition occurs at $h/J=3b$ for $T/J=0$.

We carefully investigate the phase transition from the $\sqrt{3} \times \sqrt{3}$ to the $3 \times 3$ LRO phase occurring at a finite temperature for $0.46 \lesssim h/J < 0.6$ [inset of Fig.~\ref{Fig3}(d)].
In the single spin flip, the ergodicity is practically broken near the transition temperature.
The $z$ components of the spins in the $\sqrt{3} \times \sqrt{3}$ and the $3 \times 3$ LRO phases are approximately consistent with the 1/3- and 1/9-plateau states illustrated in Figs.~\ref{Fig1}(d) and \ref{Fig1}(e), respectively. 
It is evident that a significant energy barrier exists between these ordered states at low temperatures in the single spin flip. 
We found the autocorrelation time diverging as the temperature approaches the transition point, indicating a dynamical phase transition into a glassy phase.
There is an infinite number of local minima in the glassy phase, at which the local configurations of the two ordered states are randomly arranged.
We here focus on the clusters highlighted by the orange area in Fig.~\ref{Fig1}(e) and introduce a multi-spin-flip update to avoid the dynamical transition and study the thermodynamic properties of the system.
Nevertheless, slow dynamics and the glassy phase would be of great interest in experimental studies.

\begin{figure}[t]
\centering
\includegraphics[width=\linewidth]{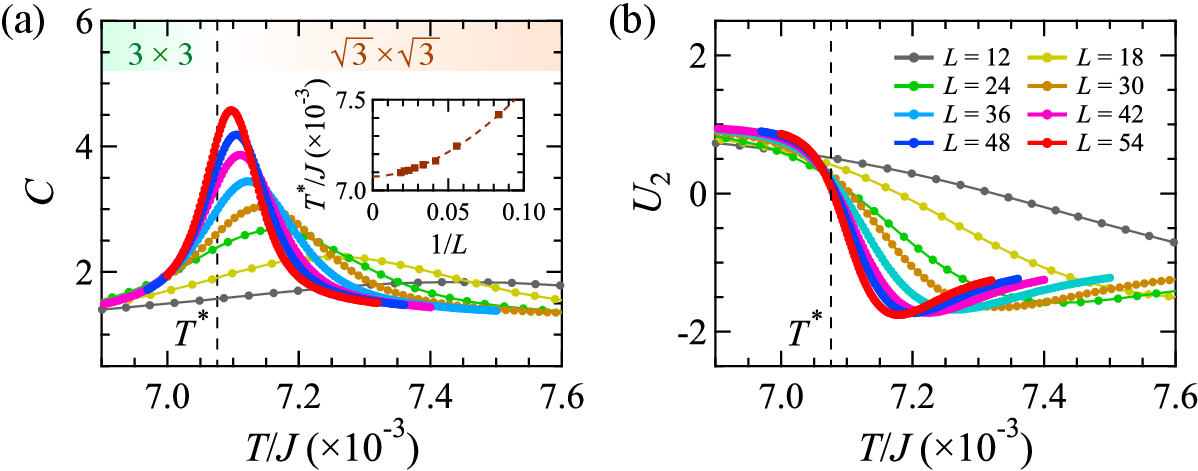}
\caption{(a) Specific heat $C$ and (b) Binder cumulant of the $3 \times 3$ order parameter $U_{2}$ as a function of temperature at $h/J=0.59$ for $b=0.2$. The inset of (a) shows the finite-size scaling of the peak position of the specific heat.}
\label{Fig4}
\end{figure}
Figure~\ref{Fig4} shows the temperature dependence of the specific heat $C$ and the Binder cumulant of the $3 \times 3$ order parameter at $h/J=0.59$ for $b=0.2$.
The Binder cumulant of an $n$-component order parameter can be defined as $U_2 = \frac{n+2}{2} \left(\frac{\langle S(\mathbf{Q})\rangle ^2}{\langle S(\mathbf{Q}) ^2\rangle} - \frac{n}{n+2} \right)$, and we here set $\mathbf{Q}=(0, \frac{4\pi}{3\sqrt{3}})$ and $n=6$ because of the three-dimensional complex vector.
Note that in $L \to \infty$, $U_2 \to 1$ and $0$ in the ordered and disordered phases, respectively.
Upon cooling, the $\mathbb{Z}_{3}$ symmetry is broken in the intermediate-temperature $\sqrt{3} \times \sqrt{3}$ LRO phase, and subsequently the remaining $\mathbb{Z}_{6}$ symmetry is further broken in the low-temperature $3 \times 3$ LRO phase.
In 2D, there are three scenarios of the $\mathbb{Z}_{6}$ symmetry breaking \cite{1977_Jos,2021_Rak}: (i) two phase transitions describing the $\mathbb{Z}_{2}$ and $\mathbb{Z}_{3}$ symmetry breaking, (ii) a two-step melting through an intermediate critical phase, and (iii) a direct first-order transition.
The specific heat shows a single peak below the $\sqrt{3} \times \sqrt{3}$ ordering temperature [Fig.~\ref{Fig4}(a)].
The Binder cumulant becomes significantly negative right above the transition temperature [Fig.~\ref{Fig4}(b)], showing no sign of a critical phase in which $U_2$ will be independent of the system size \cite{2019_Sur}.
These results presumably indicate the scenario (iii) \cite{2021_Rak}.
Using the scaling $T^*(L) - T^*(\infty) \propto L^{-\theta}$, where $T^*(L)$ is the peak position of the specific heat, we estimate $T^*/J=0.007076(4)$ and $\theta \approx 1.8$.
The exponent is reasonably consistent with the case of the first-order transition $\theta=d=2$, where $d$ is the system dimension. 
We therefore conclude that this transition is of weakly first order.

\section{Comparison with the pyrochlore system}

We here compare the effects of SLC on the kagome- and pyrochlore-lattice Heisenberg antiferromagnets.
For the pyrochlore case, the bond-phonon model, in which the SLC only produces the effective biquadratic terms $-b({\mathbf S}_{i} \cdot {\mathbf S}_{j})^2$ between the NN sites, induces a nematically ordered 2-up–2-down state at zero field \cite{2010_Sha, 2021_Aoy}.
The site-phonon model, in which the additional three-body terms ${\mathcal{H}_{\rm FN}}$ expressed by Eq.~(\ref{H_FN}) are effectively produced, lifts the macroscopic degeneracy remaining in the bond-phonon model \cite{2006_Ber, 2016_Aoy}.
For the kagome case, by contrast, the macroscopic degeneracy remains in the zero-field coplanar (collinear) ground states even in the site-phonon model because the energies of ${\mathcal{H}_{\rm FN}}$ for the spin configurations I and II (III and IV) are accidentally identical [Figs.~\ref{Fig1}(a) and \ref{Fig1}(b)].
The difference in the ground-state nature between the kagome and pyrochlore systems is due to the different ratio of the numbers of the second- and third-NN exchange paths: i.e., 1:1 in the kagome lattice, whereas 2:1 in the pyrochlore lattice.

On the other hand, common features can be seen in field-induced magnetization plateau states.
For the pyrochlore case, the 16-sublattice 3-up–1-down state (called the {\it R} state) with $m=1/2$ rather than the four-sublattice ${\mathbf q}=0$ one is stabilized in the site-phonon model \cite{2006_Ber, 2021_Aoy}.
This is because the site-phonon model favors a down-up-down spin configuration on a bent path as shown in the configuration III [Fig.~\ref{Fig1}(b)] (the ``bending rule'') rather than one on a straight path \cite{2006_Ber}.
It has been proven that the {\it R} state is a unique solution that can maximize the number of such bent paths for the pyrochlore lattice \cite{2006_Ber}.
Indeed, the {\it R} state was experimentally confirmed in the pyrochlore-based chromium spinels CdCr$_{2}$O$_{4}$ and HgCr$_{2}$O$_{4}$ \cite{2007_Mat, 2010_Mat}.
The bending rule also holds for the kagome case; the $\sqrt{3} \times \sqrt{3}$ ordered 2-up–1-down state [Fig.~\ref{Fig1}(d)] is clearly a unique solution with the maximum number of the down-up-down bent paths for $m=1/3$.
The newly found $3 \times 3$ ordered 5-up–4-down state with $m=1/9$ [Fig.~\ref{Fig1}(e)] also consists of local spin configurations with as many bent paths as possible.
This universality of the magnetization plateau nature, regardless of the dimensionality of the underlying triangular network, is an important property of the antiferromagnet described by the combination of highly symmetric geometrical frustration and SLC.

\section{Summary and perspective}
In summary, we have thoroughly investigated the effect of SLC on the magnetic properties of the classical KHAF using the microscopic magnetoelastic model.
We found a nematically ordered state at zero field and a robust 1/9-magnetization plateau in the strong SLC ($b>1/6$).
The thermodynamic properties were investigated in detail by means of the classical MC simulations.

Finally, we discuss the relevance of our results to experiments.
The zero-field ground states of KHAF compounds (in a {\it structurally perfect} kagome lattice) have been intensively studied to this time; some of them are proposed to be the 120$^{\circ}$ coplanar state stabilized by the DM interaction \cite{2000_Ita, 2017_Oku, 2020_Iid} while others a possible quantum spin liquid \cite{2015_Fu, 2016_Sun, 2021_Liu}.
In either case, the SLC does not seem to affect the magnetic states, which is consistent with our result showing that the weak SLC has a minor effect on the low-temperature magnetism at zero field.
This may be the reason why the effect of SLC has been overlooked in many studies of magnetism on the kagome lattice.
However, the SLC can play an essential role in the magnetization process.
For several KHAF compounds, the appearance of a 1/3-magnetization plateau has been confirmed by pulsed high-field magnetization measurements \cite{2019_Oku, 2020_Oku, 2011_Oku}.
Even though such observations have been attributed to thermal or quantum fluctuations in the literature, our calculation suggests that the SLC can further stabilize the magnetization plateau.

We note that most of the previously reported KHAF compounds are Cu-based $S=1/2$ quantum magnets, where SLC tends to be relatively weak because the ligand anion mediates the antiferromagnetic (AFM) exchange coupling between the NN Cu sites, and the relative strength of SLC to the exchange coupling is proportional to the square of the spin length $S^{2}$.
The parameter $b$ = 0.1–0.2 studied in the present paper is plausible in real systems with a larger spin length, such as chromium spinels \cite{2006_Ued, 2008_Koj, 2019_Gen, 2015_Kim}, where $S=3/2$ Cr$^{3+}$ ions govern the magnetism, and the AFM direct exchange interaction between the NN Cr sites is dominant.
Relevant KHAFs could be created by partial ion substitution of these pyrochlore-lattice compounds \cite{2016_Dun}.
We hope our work encourages the search for new KHAF compounds that realize our theoretical predictions.

\begin{acknowledgments}
The authors are grateful to Y. Ishii and T. Okubo for the fruitful discussions.
Some simulations were performed using computational resources of the Supercomputer Center at the Institute for Solid State Physics, the University of Tokyo. 
This work was partly supported by the JSPS KAKENHI Grants-In-Aid for Scientific Research (Grant No. 20J10988).
M.G. was supported by the JSPS through a Grant-in-Aid for JSPS Fellows.
\end{acknowledgments}

\appendix

\section{Monte Carlo updates}
We here describe the details of our MC updates.
To achieve efficient computation, we introduce a microcanonical update in addition to the conventional updates: the random update and the over-relaxation-like update \cite{2014_Shi, 2021_Aoy}. 
In the random update, a spin direction is randomly proposed for each spin, and the proposed configuration is accepted or rejected using the Metropolis algorithm for the effective model, Eq.~(\ref{H_eff}). In the over-relaxation-like update, the spin configuration $\pi$ rotated about a molecular field from the current configuration is proposed for each spin. The molecular field is calculated only taking into account the bilinear and the Zeeman terms in Eq.~(\ref{H_eff}). The proposed state is then accepted or rejected like the random update. 

In addition to these conventional updates, we implement a microcanonical update with the total energy conserved, returning back to the original Hamiltonian~(\ref{H}) with the lattice degrees of freedom. Because this Hamiltonian is quadratic in terms of the spin and the lattice displacement, we can perform perfect over-relaxation updates for spins and lattice displacements without changing the total energy \cite{1981_Adl,1996_Alo}. Specifically, the next spin configuration and displacement for site $i$ are given by
\begin{align}
    {\mathbf S}'_i &= \frac{2 \left( {\mathbf S}_i \cdot {\mathbf h}_i\right) {\mathbf h}_i }{| {\mathbf h}_i|^2} - {\mathbf S}_i,\\
    {\mathbf u}'_i &= 2 \bar{\mathbf u}_i - {\mathbf u}_i,
\end{align}
respectively, where
\begin{align}
    {\mathbf h}_i &= {\mathbf h}_{\rm ext} - \sum_{j \in N(i)} J_{ij} {\mathbf S}_j
\end{align}
with ${\mathbf h}_{\rm ext}=(0, 0, h)^t$. Note that $J_{ij}$ depends on local lattice displacements. We update spins and displacements sequentially.
To perform this microcanonical update, the lattice degrees of freedom can be restored generating a displacement $\mathbf{u}_i$ for each site from the Gaussian distribution whose mean is $\bar{\mathbf u}_i$, and the variance is $T/c$, where $T$ is the temperature. 

The random or the over-relaxation-like update step alternately follows several microcanonical update steps in our simulation. 
The single MC step in our simulation is composed of $N=3L^2$ local updates for spins and displacements sweeping all sites sequentially in one of the updates mentioned above: the random, the over-relaxation-like, or the microcanonical update. 
More than $2^{24}$ MC steps were run, and the latter half was used to calculate the averages of the physical quantities.

The microcanonical update we introduce significantly reduces the autocorrelation time. We calculated the integrated autocorrelation time $\tau_{\rm int}$ of the energy for $b=0.2$, $h/J=0$, $T/J=0.04$, and $L=12$ and estimated $\tau_{\rm int}\approx 2.4 \times 10^3$ and $1.0 \times 10^5$ with and without the microcanonical update, respectively. Thus, the sampling efficiency of our approach is approximately $42$ times as high as that of the previous approach. 

As discussed in Sec.~\ref{sec:in-h}, the ergodicity in the local updates is practically broken near the finite-temperature transition from the $\sqrt{3} \times \sqrt{3}$ to the $3 \times 3$ LRO phase. We found the autocorrelation time diverging as the temperature approaches the transition point, indicating a dynamical transition into a glassy phase in which the local configurations of the two ordered states are randomly arranged.
In addition to the local updates, we further introduce a multi-spin-flip update to avoid the dynamical transition, which allows us to study the thermodynamic properties of the system. 
Flipping the spins highlighted by the orange area in Fig.~\ref{Fig1}(e) connects the 1/3- and 1/9-plateau states. 
We focus on a nine-spin cluster consisting of the highlighted neighboring spins around a hexagon of the kagome lattice as a minimal cluster. 
We sequentially choose a nine-spin cluster and perform the $z$-component flip of the cluster, using the Metropolis algorithm. Although the acceptance probability of the multispin update may be fractional, this update process significantly helps the system escape from local minima and thermalize. We successfully calculated the thermodynamic quantities and obtained the phase boundary thanks to the introduced microcanonical and the multispin updates. 

\section{Dynamic temperature optimization for efficient replica exchange}
We used the replica-exchange MC \cite{1996_Huk} to reduce the autocorrelation further. The temperature exchange process followed each MC step explained above.
Because the original model, Eq.~(\ref{H}), has the additional energy fluctuations of the lattice displacements, the effective model, Eq.~(\ref{H_eff}), is preferred to the original model for increasing the exchange probability. We consider the original model when performing the microcanonical update; we use the effective model for the other updates, including the replica-exchange process. The number of the temperatures in the replica-exchange MC was typically several hundreds in the present simulations.

We here discuss the optimization of the temperature set for efficient computation. It has been argued that the exchange probability is desired to be independent of the temperature \cite{2002_Kof}. In accordance with this criterion, we aim at the flat distribution of the exchange probability given the temperature bounds and the number of temperatures. We dynamically optimize the set of inverse temperatures $\{ \beta_i\}$ in a manner similar to the stochastic approximation \cite{2015_Yas}. Specifically, we set the distance of the adjacent inverse temperatures larger if the exchange process is accepted and smaller otherwise. The modification factor gradually decreases with time, that is, the number of preceding exchange steps. Given the number of temperatures $n$ and the maximum and the minimum inverse temperatures $\beta_{\rm max}$ and $\beta_{\rm min}$, respectively, the optimization procedure is described as follows:
\begin{itemize}
    \item  Set an initial set of $\beta$, $\beta_{\rm min} \equiv \beta_1 < \beta_2 < \cdots < \beta_n \equiv \beta_{\rm max}$ with $\Delta \beta_i=\beta_{i+1} - \beta_i$.
    \item  For each exchange step $t=1,2,\dots,t_{\rm opt}$, exchange the replicas with $\beta_i$ and $\beta_{i+1}$ with a probability $p=\min \{ 1, \exp (\Delta \beta_i \Delta E_i) \}$, where $\Delta E_i = E_{i+1} - E_i$, and $E_i$ and $E_{i+1}$ are the energies of the corresponding replicas. The replica index $i$ runs over odd integers when $t$ is odd and even integers when $t$ is even. If the exchange is accepted, calculate
\begin{equation}
    \Delta \beta_i' = \Delta \beta_i + a \frac{\beta_{\rm max} - \beta_{\rm min}}{t};
\end{equation}
otherwise
\begin{equation}
    \Delta \beta_i' = \Delta \beta_i - \min \left\{ \frac{\Delta \beta_i}{2}, a \frac{\beta_{\rm max} - \beta_{\rm min}}{t} \right\},
\end{equation}
where $a$ is a parameter.
For even $i$ at odd $t$ and odd $i$ at even $t$, $\Delta \beta_i' = \Delta \beta_i$.
After calculating all $\Delta \beta_i'$, update $\Delta \beta_i$ by normalizing $\Delta \beta_i'$, that is,
\begin{equation}
\Delta \beta_i = \Delta \beta_i' \frac{\beta_{\rm max} - \beta_{\rm min}}{\sum_j \Delta \beta_j'},
\end{equation}
and set $\beta_i = \beta_1 + \sum_{j=1}^{i-1} \Delta \beta_j$ for $1 < i < n$.
\end{itemize}
In the present simulations, we set $a=10$ and the optimization period $t_{\rm opt}$ to be half of the total number of MC steps in the thermalization process. After the temperature optimization, we fixed the temperature set and calculated the mean acceptance probability for each temperature during the sampling. The resultant probability was successfully almost independent of the temperature: the maximum deviation of the acceptance probability from the average was typically only a few percent of the probability averaged over the temperatures. 

\section{Stochastic approximation}
We used the stochastic approximation (SA) \cite{2015_Yas} to obtain the phase boundaries on which the continuous or the BKT transition occurs. The SA is a useful approach to finding a root of a function in the presence of stochastic errors in the function evaluation. A control parameter, which is the temperature in our application, is dynamically optimized to find a solution. The feedback factor to the control parameter gradually decreases with time, or the number of steps, which makes the estimation robust to the stochastic error.

To locate the phase transition point to the $\sqrt{3} \times \sqrt{3}$ LRO phase, we optimize the temperature for each system size $L$ to satisfy the condition
\begin{equation}
\label{sa_cond}
    f(T) \equiv R - \frac{\xi}{L} = 0,
\end{equation}
where $T$ is the temperature, $R>0$ is a parameter, and $\xi$ is the corresponding correlation length, which can be calculated by the second (or the higher-order) moment method \cite{2015_Suw}:
\begin{align}
    \xi &= \frac{1}{|{\boldsymbol \delta}|} \sqrt{\frac{S({\mathbf Q})}{S({\mathbf Q}+{\boldsymbol \delta})}-1},\\
    S(\mathbf{Q}) &= \frac{1}{N} \left\langle \left| \sum_j \mathbf{O}_j e^{i \mathbf{Q} \cdot \mathbf{r}_j} \right|^2 \right\rangle, 
\end{align}
with ${\mathbf O}_j={\mathbf S}_j$ or $\bar{\mathbf u}_j$ and $\mathbf{Q}$ being the ordering wave vector. The choice of $R$ is arbitrary in principle, but setting $R$ close to the critical amplitude is practical \cite{2015_Yas}. We set $R=0.5$ in the present simulations. To detect the $\sqrt{3} \times \sqrt{3}$ order in the kagome lattice, we used $\mathbf{Q}=(\frac{4\pi}{3}, 0)$ and $\boldsymbol{\delta} = (\frac{2\pi}{L}, \frac{2\pi}{\sqrt{3}L})$ for each $L$. In practice, we found more efficient using ${\mathbf O}_j={\mathbf S}_j$ for small $h$ and ${\mathbf O}_j=\bar{\mathbf u}_j$ for large $h$ because of the smaller finite-size corrections. 

For $t=1,2,\dots,t_{\rm SA}$, the temperature is dynamically updated during the simulation:
\begin{equation}
    T^{(t+1)}=T^{(t)} - \frac{p}{t} f \left( T^{(t)} \right),
\end{equation}
where $T^{(t)}$ is the temperature in the $t$th step of the SA, and $p$ is a parameter.
We roughly set $p \sim \frac{1}{f'(T_{\rm c})}$ to achieve the fastest convergence, where $f'(T_{\rm c})$ is the derivative at the solution (critical point) \cite{2015_Yas}. An optimized temperature $T_{\rm c}(L)$ for each $L$ was obtained such that Eq.~(\ref{sa_cond}) was approximately satisfied. We extrapolated the phase transition point from the optimized temperatures $T_{\rm c}(L)$, using the asymptotic scaling form
\begin{equation}
    T_{\rm c}(L) - T_{\rm c}(\infty) \propto L^{-1/\nu},
\end{equation}
where $\nu$ is the critical exponent of the correlation length.

The condition to satisfy in the SA can be selected for each phase transition. For detecting the BKT transition of the integer charged vortex, we used the Nelson-Kosterlitz formula $\rho_s(T_{\rm BKT})=\frac{2}{\pi v^2}T_{\rm BKT}$ \cite{1977_Nel} with $v=1$, where $\rho_s$ is the spin stiffness. Thus, we set
\begin{equation}
    f(T)=\frac{2}{\pi}T - \rho_s.
\end{equation}
The asymptotic scaling of the optimized transition temperature $T_{\rm BKT}(L)$ is given by
\begin{equation}
        T_{\rm BKT}(L) - T_{\rm BKT}(\infty) \propto \left( \log L \right)^{-2}
\end{equation}
because the correlation length exponentially diverges, $\xi \sim e^{\frac{A}{\sqrt{T-T_{\rm BKT}}}}$ with a constant $A$.

We typically ran $2^{15}$ MC steps for each SA step to calculate the physical quantity and set $t_{\rm SA} \sim 10^3$. We averaged the obtained $T_{\rm c}(L)$ and $T_{\rm BKT}(L)$ over more than ten independent SA simulations.

\end{document}